\documentclass[preprint,prb,showpacs]{revtex4}
\usepackage{psfrag}
\usepackage{amssymb}
\usepackage{indentfirst}
\usepackage[dvips]{graphicx}
\usepackage{graphics}

\begin{document}
\title{Interface heat transfer between crossing carbon nanotubes, and the thermal conductivity of nanotube pellets
}
\author{Yann Chalopin$^1$}
\author{Sebastian Volz$^1$}
\author{Natalio Mingo$^{2,3}$\footnote{Corresponding author, e-mail: natalio.mingo@cea.fr}}

\affiliation{$^1$Laboratoire d'Energ\'etique Mol\'eculaire et 
Macroscopique, Combustion\\
CNRS UPR 288, Ecole Centrale Paris,
Grande 
Voie des Vignes, F-92295 Ch\^atenay-Malabry cedex, France\\
$^2$LITEN, CEA-Grenoble, 17 rue des Martyrs, 38054 Grenoble Cedex 9, France \\
$^3$Jack Baskin School of Engineering, UCSC, Santa Cruz, California 95064, USA} 

\linespread{1}
\begin{abstract}
We theoretically compute the interface thermal resistance between crossing single walled carbon nanotubes of various chiralities, using an atomistic
Green's function approach with semi-empirical potentials. The results are then used to model the
thermal conductivity of three dimensional nanotube pellets in vacuum. For an average nanotube length of 1 $\mu$m, the model yields an upper bound for the thermal conductivity of densely compacted pellets,
of the order of a few W/m-K. This is in striking contrast with the ultra-high thermal conductivity reported on individually suspended nanotubes. The results suggest
that nanotube pellets might have an application as thermal insulators.
\end{abstract}
\maketitle

\section{Introduction.}

Tremendous advancements in nanoscale heat transfer provide new alternatives to managing phonon transport inside crystalline nanosized devices. As the typical lenghtscale of systems reach those of the dominant phonon wavelengths, the transport of heat occurs in a coherent regime. Recently, various approaches have been developed to reconcile the wave behaviour of phonons with the diffusive and non-diffusive (ballistic) processes occuring inside confined solid state devices \cite{che2,che1,cro1,min1}. Thermal management in short-scaled systems cannot be carried out by taking into account the classical Fourier law. Impurities, defects, Umklapp processes as well as roughness play a key role in phonon scattering: the flow of heat drastically differs from that at the macroscale \cite{cah1,gla1,dam1,lu1}.

For the past decade, thermal transport inside carbon nanotubes as well as inside semiconductor nanowires have been intensively investigated for their ballistic transport properties. Experimental mesurements on individual CNT as well as theoretical predictions confirmed a very high thermal conductivity \cite{yu1,kim01,ber1,pad1,hon1}. Nonetheless, even if single-walled CNT exhibit a ballistic regime in conducting heat, recent measurements showed that conductivity gets drastically reduced in composites \cite{hux1,xie1,Nan1,yan1}.
 This is explained by the contribution of the interfacial contact resistances. As the percolation treshold is reached by incorporating CNT, an interconnection pattern is formed and thus, each CNT gets connected to a network structure through different contacts depending on the mass concentration. These contacts are responsible for reducing the heat flux in the material. As the structure gets compacted, interface effects become predominant.

A few works have reported on the impact of interface resistances from an atomistic approach \cite{zho1,Carlborg,Kumar,wen1,Saha,WZhang,tae1}. Ref.~\onlinecite{zho1} performed molecular dynamic simulations to predict the contact  thermal conductance dependance due to the length of the overlap, and the spacing between two aligned overlaping CNT. This work showed that reducing the contact area increased the contact resistance.  The predominance of the interface contact resistance over the nanowire/nanotube intrinsic resistance has also been reported from a theoretical point of view for the case of a planar substrate connected to a silicon nanowire at low temperatures \cite{cha1,pra2}. The physical mechanism of this very low conductance is explained in terms of a phonon density of states mismatch at the junction.

Relevant numerical techniques used to retrieve thermal properties are mostly molecular dynamics (MD), the Boltzmann transport equation, and atomistic Green's functions (AGF). 
MD is very well suited to solve high temperature problems because anharmonicities are usually included in the interatomic potentials,  the many body interaction are by the way fully captured.
When harmonic scattering (due to interfaces or defects for example) is predominant, thermal transport is best described by atomistic Green's functions \cite{min2}. Pertubative treatments can also be used to capture high temperature anharmonic phonon scattering effects \cite{min3}.

In this paper, we first use the Green's function formulation of heat transport to obtain the phonon partial-transmission functions at the junction of two crossing overlaping CNT (CCNT) for different chiralities.
Afterwards, we calculate the partial conductances at the junction and their dependence with CNT chirality. 
We then present a simple model involving the previously calculated interface resistances, which predicts the thermal dependence of CNT composites conductivity as a function of the CNT mass density. We discuss the dependence of the thermal conductivity on the various physical parameters characterizing the pellet. An interesting result stemming from the calculation is that the thermal conductivity has an upper bound, for dense pellets, which does not depend on the diameter of the nanotubes in the composite. This upper bound is several orders of magnitude smaller than the thermal conductivity of individual carbon nanotubes. We finally discuss the implications of these findings for novel technological applications.

\section{Phonon transport at the junction between crossing nanotubes.}

\begin{figure}[htbp]
\begin{center}
\includegraphics[scale=0.27]{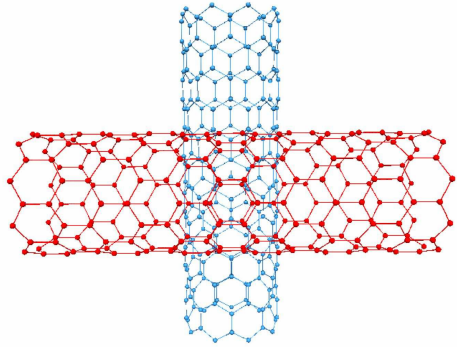}
\caption{Typical crossing configuration of two carbone-nanoubes in a nematic (compressed/aligned and non-isotropic) phase. }
\label{fig:conf}
\end{center}
\end{figure}

A typical crossing nanotube configuration is shown in Fig. \ref{fig:conf}. 
The theory of harmonic interface thermal conductance calculation using Green's functions was developed in Ref.~\onlinecite{min2}. It is straghtforward to
generalize the theory in order to compute partial phonon transmissions between any two of the four nanotube edges. The case that we are interested in is the one where 
each nanotube is kept at a given temperature, equal on both edges. In such case, the thermal current across the junction is given by
\begin{eqnarray}
J=\int{\hbar\omega {\cal T}(\omega)(f_u-f_l)d\omega/2\pi},
\end{eqnarray}
where $f_{u(l)}$ is the Bose-Einstein distribution corresponding to the temperature of the upper (lower) nanotube in the junction.
The thermal conductance across this junction is obtained as $\sigma=dJ/dT$, and given by
\begin{equation}
\sigma  = \int\limits_0^{\omega _{\max } } {{\cal T} (\omega )\frac{\partial }{{\partial T}}\left( {\frac{1}{{e^{\hbar \omega /k_B T}  - 1}}} \right)\hbar \omega \frac{{d\omega }}{{2\pi }}} 
\label{equcond}
\end{equation}
Ways to compute the transmission function have been explained in previous publications. Here we employed a three-region formula, given by
\begin{eqnarray}\label{threereg}
\mathcal{T}(\omega)=Tr(\mathbf{G}_{32}\Gamma_2[\mathbf{G}^+]_{23}\Gamma_3),
\label{equ.T}
\end{eqnarray}
where $\Gamma_3\equiv \mathbf{k}_{34}(\mathbf{g}_{44}-\mathbf{g}_{44}^+)\mathbf{k}_{43}$.
Other ways of obtaining the transmission are also possible, like the two-region formula, for example, and the results are equivalent in all cases \cite{MingoChapter}.

In order to compute the Green functions, a proper description of the system's atomic configuration and interatomic interactions is necessary. We have
employed the Brenner potential to compute energies and forces between atoms belonging to the same nanotube \cite{bre1}. Interactions between atoms in different nanotubes 
are not given by the Brenner potential. These are weaker interactions, and it is customary to describe them by a Lennard-Jones potential \cite{zho1}. Several parametrizations
can be found in the literature. We have opted for the one used by Zhong and Lukes \cite{zho1}, but results are not expected to change much if another reasonable parametrization
were used. The parameters are $\epsilon=4.41meV$ and $\sigma=0.228nm$

The stable atomic arrangements were obtained by energy minimization. A large supercell periodic array of perpendicularly crossing nanotubes was constructed for this purpose.
The atoms in the last unit cells at the four edges of the two crossing nanotubes in the supercell were kept in a rigid structure corresponding to the relaxed isolated nanotube
lattice. This is needed in order to seamlessly join the crossing part to the four semi-infinite contacts afterwards, for the Green's function calculation.
The large size of the supercell ensures that relaxation is achieved within the crossing region.
The two nanotubes' relative distance, crossing
position, and angle around their axes, were varied until the lowest energy minimum was found. Afterwards, the interatomic force constants of the
system were numerically computed by finite differences between the forces upon small (0.00001 \AA) displacements of each degree of freedom.
Force constant matrices were independently calculated for the isolated nanotubes, to be used for the projection of the semi-infinite contacts.

The self-energies corresponding to the four semi-infinite contacts were computed using a decimation technique \cite{gui1}. Then, the total Green's function was obtained from the
previously calculated force constants for the crossing region, and the phonon transmissions were evaluated using Eqn. \ref{equ.T}.

\begin{figure}
\begin{center}
\includegraphics[scale=0.32]{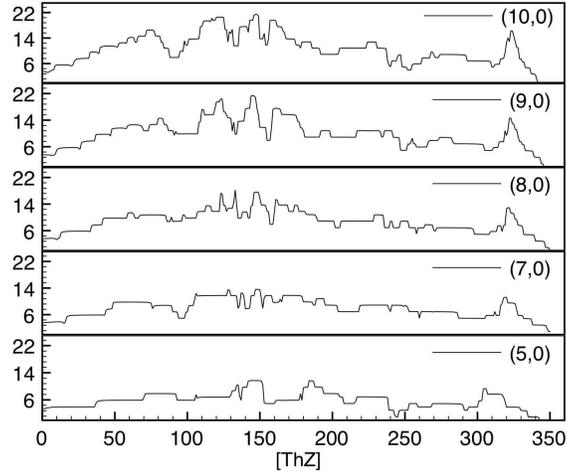}
\caption{Phonon Transmission functions through the same nanotube in the presence of the crossing one.}
\label{fig:TransSame}
\end{center}
\end{figure}

\begin{figure}[htbp]
\begin{center}
\includegraphics[scale=0.32]{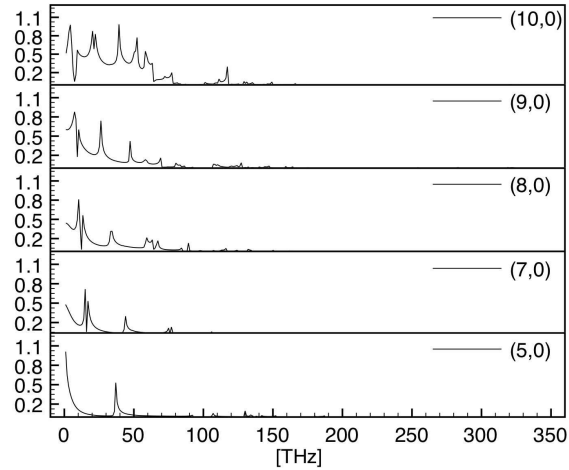}
\caption{Phonon Partial Transmission functions between two crossing nanotubes of identical chirality (Zigzag cases).}
\label{fig:Trans}
\end{center}
\end{figure}

\begin{figure}[htbp]
\begin{center}
\includegraphics[scale=0.32]{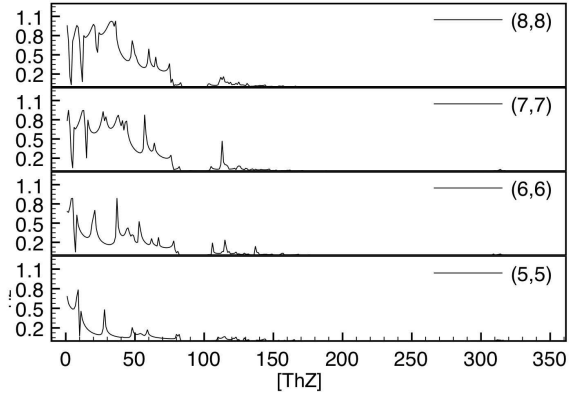}
\caption{Phonon Partial Transmission functions between two crossing nanotubes of identical chirality (Armchair cases).}
\label{fig:TransArm}
\end{center}
\end{figure}

We start by looking at the phonon transmission inside one individual CNT, in the presence of another CNT crossing it. Fig.~\ref{fig:TransSame} shows that the effect of the crossing is very small: the transmission function is very slightly reduced with respect to the integer transmission values for an isolated nanotube.
In contrast, comparing Fig.~\ref{fig:TransSame} with Figs.~\ref{fig:Trans} and \ref{fig:TransArm} shows that the phonon transmission between two different CNT is much lower than the one on the same CNT. Roughly speaking, this means that each nanotube can be considered to be at a constant temperature, and temperature differences exist only at the interfaces between nanotubes.

\begin{figure}[htbp]
\begin{center}
\includegraphics[scale=0.3]{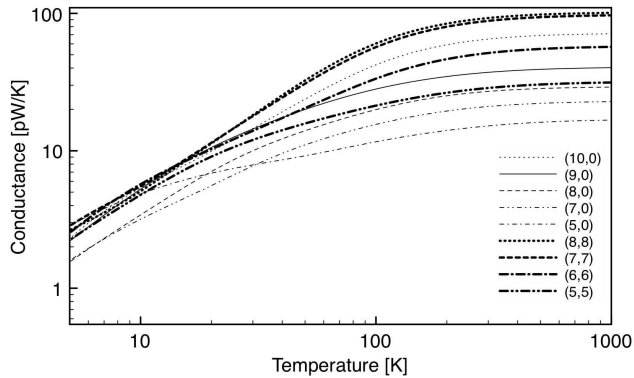}
\caption{Interface conductance for (5,5),(6,6),(7,7),(8,8) - (5,0),(7,0),(8,0),(9,0) and (10,0) chiralities at 300K. Crossing configurations are set for two identical CCNT. }
\label{fig:cond}
\end{center}
\end{figure}

Figs.~\ref{fig:Trans} and \ref{fig:TransArm} also reveal that at low frequencies, transmission is finite (non-zero.) This means that quantization effects will become observable at very low temperature, in the form of a linear temperature depencence of the thermal conductance. The interface thermal conductances evaluated from the transmission functions are shown in Fig.~\ref{fig:cond}, as a function of temperature. Due to the strong suppression of the interface transmission above a few tens of THz, the interface thermal conductance temperature dependence is very weak above 100K. This contrasts with the isolated nanotube thermal conductance, which increases noticeably between 100K and 300K \cite{min05}. We will now use these computed thermal conductance values to estimate the thermal conductivity of CNT pellets, in the next section.

\section{The thermal conductivity of compact CNT pellets}

\begin{figure}[htbp]
\begin{center}
\includegraphics[scale=0.45]{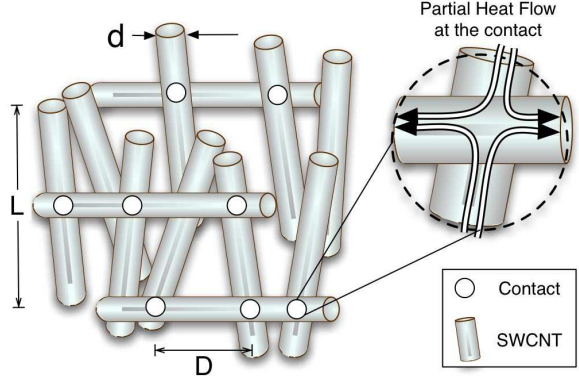}
\caption{Length parameters defining the CNT pellet :  $l=$ nanotube length, $d=$ nanotube diameter and $D\sim$ segment length between junctions}
\label{fig:pelet}
\end{center}
\end{figure}

A pellet is formed by a randomly arranged network of finite length nanotubes, intercrossing at various
points along their lengths. As sketched in Fig.~\ref{fig:pelet}, there are three length parameters defining the pellet: $l=$ nanotube length, $d=$ nanotube diameter, and $1/D^3=$ density of junctions, with $D\sim$ segment length between junctions. As we have seen in the section above, placing a junction does not appreciably diminish the
phonon transmission along the same nanotube. Thus, to a good approximation (even if the nanotubes are micrometers long, and
have junctions every 10 nanometers) the temperature on each nanotube can be considered constant. For a macroscopic pellet, on 
average, the temperature of a nanotube whose center of mass is at ${\vec x}_i$ will correspond to $T_i=T_{cold}-\vec{\nabla} T\cdot {\vec x}_i$,
where $T_{cold}$ is the temperature of the colder side.

The heat flux is thus determined by the amount of heat traversing each individual junction across the pellet. Let us assume that
the density of junctions per unit volume is $1/D^3$, i.e. the average distance between junctions is roughly $\sim D$. The thermal current
per unit cross section is then $J/S = \sigma {\langle |\Delta T_{junct}|\rangle}/ D^2$,
where $\sigma$ is the junction's thermal conductance computed in the previous section, and ${\langle |\Delta T_{junct}|\rangle}$ is the absolute temperature difference at the two sides of the
junction, averaged throughout the pellet.

In order to evaluate ${\langle |\Delta T_{junct}|\rangle}$, we average over all
possible orientations $\Omega_1$, $\Omega_2$ of the two nanotubes, and contact positions $x_1$, $x_2$ measured along the two nanotubes. Denoting the nanotube length by $l$, we have
\begin{eqnarray}
\langle |T_1-T_2|\rangle/\nabla T = 
{1\over l^2}\int_{-l/2}^{l/2}\int_{-l/2}^{l/2}dx_1 dx_2
\int\int{d\Omega_1 d\Omega_2\over (4\pi)^2}|z_1-z_2| \\
=l{1\over 4}\int_{-l}^{l}d\tilde x_1\int_{-1}^{1}d\tilde x_2
\int_{-1}^{1}{d(\cos\theta_1)\over 2\pi}\int_{-1}^{1}{d(\cos\theta_2)\over 2\pi}
|\tilde x_1\cos\theta_1-\tilde x_2\cos\theta_2|\\
=l{5\over 27}\simeq 0.18 l.
\end{eqnarray}

Combining this result with those in the previous paragraph yields the pellet's thermal conductivity,
\begin{eqnarray}
\kappa\equiv{J\over S \nabla T}\simeq {0.18 l \over D^2}\sigma.
\label{kappaformula}
\end{eqnarray}

Since the average distance between junctions cannot be smaller than the nanotube diameter, there exists
an upper bound to the pellet's thermal conductivity upon compaction. In principle, this upper limit
is larger the longer the nanotubes are. However, if the nanotubes are so long that they are effectively multiply bent
in a noodle fashion, $l$ in the equations is no longer the total nanotube length, but rather twice the main curvature radius of the nanotubes. It is possible to rewrite the previous model assuming that the nanotubes are closed circumpherences, yielding essentially the same results.

An important fact that stems from the previous equation is that the thermal conductivity of the most highly compacted pellets should be roughly independent of the diameter of the nanotubes \cite{PrivateCommun}. To see this, it is enough to realize that the junction's thermal conductance is roughly proportional to the contact area, which in turn is proportional to the square of the nanotube diameter, $d^2$. Since the minimum possible distance between junctions $D$ is also proportional to $d$, the result in Eqn.~\ref{kappaformula} is independent of $d$. Obviously, this is only true at sufficiently high temperatures. For low temperatures the thermal conductance enters the quantized regime, and it no longer scales like the contact area (see discussion of Fig.~\ref{fig:condlimit} below.)

It is convenient to express the pellet's thermal conductivity in terms of its density, so that results can be directly compared with future experiments. Realizing that there are two segments per junction, the pellet's density is
\begin{eqnarray}
\rho\simeq 2 {\pi d \rho_{graphene}\over D^2},
\end{eqnarray}
where $\rho_{graphene}=7.6\times 10^{-7}$ Kg/m$^2$ is the surface mass density of graphene.
The thermal conductivity is thus linearly proportional to the pellet's density, as
\begin{eqnarray}
\kappa\simeq \sigma {0.18 l\over 2 \pi d\rho_{graphene}}\rho.
\end{eqnarray}
The high compaction density limit, assuming again that $D\sim 2d$, is 
$\rho_c\sim{\pi \rho_{graphene}\over 2d}$.

\begin{figure}[htbp]
\begin{center}
\includegraphics[scale=0.32]{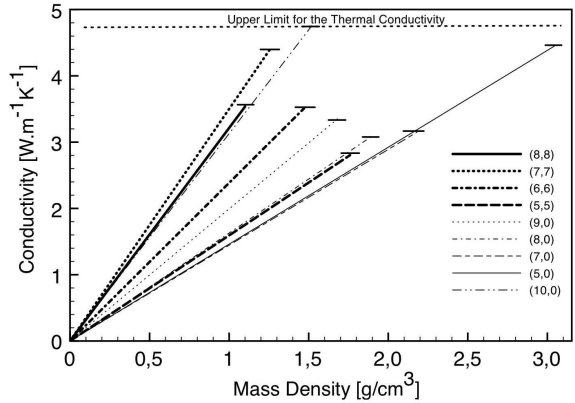}
\caption{Thermal conductivity as a function of CNT mass density for (5,5), (6,6), (7,7), (8,8), (5,0), (7,0), (8,0), (9,0) and (10,0) chiralities at 300K. Curves end-up when high density limit is reached (horizontal lines). }
\label{fig:condZig}
\end{center}
\end{figure}

To illustrate these results we have considered a typical nanotube length of $1\mu$m. Fig.~\ref{fig:condZig} shows the pellet's thermal conductivity as a function of its density, for different nanotube chiralities. A real composite will contain a range of chiralities and diameters, yielding an average of the individual results shown. Each of the lines shown is terminated at the highest possible density for nanotubes of that diameter. All the thermal conductivities in the dense limit are of the same order, and smaller than $\sim 5$ W/m-K. This very low upper bound for the thermal conductivity of the pellets contrasts with the high thermal conductivities (600 times larger) measured in isolated carbon nanotubes, which are comparable to that of graphite \cite{kim01}.

Figure \ref{fig:condlimit} shows the thermal conductivity in the high density limit, as a function of nanotube diameter, for various temperatures. Results at 300K and 1000K are roughly independent of d, whereas at 
100K the results clearly decrease with increasing d. This is a result of thermal conductance quantization effects becoming important. In the low temperature limit, the thermal conductance across the junction is given by the zero frequency value of the transmission function times the quantum of thermal conductance. As shown in section II, the transmission function is always of the order of $\sim 1$ in the low frequency limit, regardless of diameter. This results in an interface thermal conductance that does not strongly depend on diameter at low temperature, which combined with Eqn.~\ref{kappaformula} results in the observed decrease as d increases.

 \begin{figure}[htbp]
\begin{center}
\includegraphics[scale=0.5]{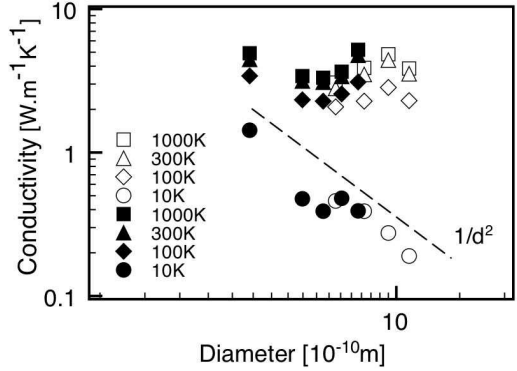}
\caption{Thermal conductivity in the high density limit as a function of the CNT diameters for ZigZag configuration (white symbol) and Armchair configuration (black symbol). Temperature is taken at 10K (circles), 300K (triangles) and 1000K (squares)  }
\label{fig:condlimit}
\end{center}
\end{figure}

This suggests that, by carefully choosing the size of the nanotubes, it might be possible to tailor the high to low temperature thermal conductivity ratio of the pellets, with possible interesting applications in thermal switches and sensors. Conversely, it would be conceivable to use the measured temperature dependent thermal conductivity in order to determine the diameter of the pellet's constituent nanotubes.

The very low thermal conductivity values of even the densest pellets suggest that there is room for carbon nanotube based materials in thermal insulation applications. Until recent times, many efforts have been devoted to trying to use nanotubes for enhancing materials' thermal conductivity. Given the inherently interesting structural and chemical properties of CNT's, low thermal conductivity carbon nanotube materials may be a still not well explored yet attractive subject of future research.

\section{Conclusions}

We have computed the interface thermal conductance between crossing carbon nanotubes of various chiralities, using an atomistic Green's function formalism with interatomic force fields. We found that interfacial heat flow is mostly carried by low frequency phonons, while phonon transmission above a few tens of THz is strongly suppressed. As a result, the interface thermal conductance depends weakly on temperature above 100K. We have then developed a model of thermal conductivity for CNT pellets that incorporates the calculated interface thermal conductance values. Using this model, we have found 
that the thermal conductivity of densely compacted pellets is severly limited by an upper bound several orders of magnitude lower than the thermal conductivity of isolated CNT's or graphite.
This suggests the possible applicability of CNT pellets in thermal insulation applications.

\section*{Acknowledgments}
We are grateful to P. Keblinski for helpful discussions.
We acknowledge O. Bourgeois from Institute Neel at Grenoble for important logistic support during the first phase of this work. Most of this work
was performed during two stays of Y. Ch. at CEA-Grenoble.

\small
\linespread{1}

 \end{document}